# ON ELECTROMAGNETIC WAVE INTERACTION WITH DENSE RESONANT ATOM MEDIUM

V. Danilov

ORNL, Oak Ridge, TN 37831, U.S.A.

*Abstract*

The excitation of atomic levels due to interaction with electromagnetic waves became of interest in accelerator physics in relation to high efficiency charge exchange injection into rings for high beam power applications. Usually, the beam density is so small that its influence on the wave is completely neglected. Here we consider the case of dense beams - the beam dimensions are large as compared to light reflection length. This paper shows that the waves can be trapped in the medium under these conditions. Moreover, the atoms with induced dipole moments start to interact strongly with each other, leading to possibility to create some atomic patterns when the medium is relatively cold.

## INTRODUCTION

It was shown in [1], that dense atom medium can trap the electromagnetic waves in resonance with atomic transitions and act like superconducting loss-free cavity for the waves. If the electromagnetic wave intensity becomes large, it applies a substantial pressure on the medium. This paper shows that one of the forces to counteract the field pressure can be the surface tension force that appears due to induced dipole moments of the resonant atoms. We present brief estimations for the energy of dipole interaction as a function of the wave field, and show that for the small temperature (1000K or less) the atoms may form the string-like or other structures. Therefore, the trapped field transforms the medium into liquid and the liquid form may contain the field inside due to induced surface tension.

## BASIC PHYSICS

Imagine now two atoms close to each other in the field of electromagnetic wave. For simplicity, we deal with hydrogen atoms, but it is generally true for all atoms or molecules having zero dipole moment in their ground state. The Schrödinger equation for two electrons reads:

$$i\hbar \frac{\partial}{\partial t}\psi = -\frac{\hbar^2}{2m}(\Delta_1 + \Delta_2)\psi + U(r_1, r_2)\psi, \quad (1)$$

with the potential function equal to

$$U(r_1, r_2) = eE_0(z_1 + z_2)\cos(\omega t) + \frac{e^2(\vec{r}_1\vec{r}_2 - 3(\vec{r}_1\vec{n})(\vec{r}_2\vec{n}))}{r^3} - \frac{e^2}{r_1} - \frac{e^2}{r_2},$$

where indexes 1, 2 related to the first and second electron, respectively, $\vec{r}$ is the vector from the first to the second nuclei, $\vec{n} = \frac{\vec{r}}{r}$, and $\vec{r}_1$ and $\vec{r}_2$ are vectors connecting nuclei with corresponding electrons. We assume the wave functions of the atoms don't overlap and we seek the solution in the following form:

$$\psi = C_1\psi_1(r_1)\psi_1(r_2)e^{-2iE_1t/\hbar} + C_2\psi_2(r_1)\psi_2(r_2)e^{-2iE_2t/\hbar} + (C_3\psi_1(r_1)\psi_2(r_2) + C_4\psi_2(r_1)\psi_1(r_2))e^{-i(E_1+E_2)t/\hbar},$$

where $\psi_1, \psi_2$ are the eigenfunctions and $E_1, E_2$ are the energy levels of unperturbed ground and upper states, respectively. After substituting to (1) one yields:

$$\dot{C}_1 = \frac{i\mu_{12}E_0}{2\hbar}(C_3 + C_4)e^{i\Delta t} - \frac{ia_1}{\hbar r^6}C_1,$$

$$\dot{C}_2 = \frac{i\mu_{21}E_0}{2\hbar}(C_3 + C_4)e^{-i\Delta t} - \frac{ia_2}{\hbar r^6}C_2,$$

$$\dot{C}_3 = \frac{i\mu_{21}E_0}{2\hbar}C_1e^{-i\Delta t} + \frac{i\mu_{12}E_0}{2\hbar}C_2e^{i\Delta t} - C_4\frac{i2|\mu_{12}|^2}{\hbar r^3}, \quad (2)$$

$$\dot{C}_4 = \frac{i\mu_{21}E_0}{2\hbar}C_1e^{-i\Delta t} + \frac{i\mu_{12}E_0}{2\hbar}C_2e^{i\Delta t} - C_3\frac{i2|\mu_{12}|^2}{\hbar r^3},$$

where $\Delta = \omega - \omega_0$, the electric field has the form $E = E_0\cos\omega t$, $\mu_{12} = \mu_{21}^* = -\int d^3r\, u_1^*(\vec{r})ezu_2(\vec{r})$ (assuming the light is polarized in the direction $z$, parallel with the atomic plane), and $u_1$ and $u_2$ are the normalized wave functions of the lower and the upper excited states, respectively. In the case of hydrogen, the lower level has primary quantum number n=1. The upper level primary quantum number is determined by the resonant condition, its angular momentum l=1, and the projection of angular momentum on the z axis m=0. For this case $\mu = \mu_{1n}$ is real and we omit its subscripts for simplicity. In addition to energy levels, we added second term of atom-atom interactions; the constants $a_1$ and $a_2$ can be found in e.g. [2].

Now the idea is to find stationary solutions, calculate the energy of dipole interaction and find its minimum - we suppose the atoms with low temperatures settle around this minimum. First thing to notice is that there are four eigenmodes for this system of equations, and if $C_3, C_4 \propto e^{i\delta\omega t}$, then $C_1 \propto e^{i(\delta\omega+\Delta)t}$ and

$C_2 \propto e^{i(\delta\omega-\Delta)t}$. We are interested in only symmetric eigenmodes $C_3 = C_4$. Thus we eliminate one (e.g., last) of the equations. The three remaining eigenmodes have two energy levels close to transition energy and one separated from them by energy of dipole interactions.

Assuming $\delta\omega, \frac{\mu_{12}E_0}{\hbar} \ll \frac{\mu_{12}^2}{\hbar r^3}$, we exclude one more eigenmodes with frequency far from the transition one.

One gets $C_3 = C_4 = \frac{E_0 r^3}{4\mu}(C_1 + C_2)$. Finally, the equation for two remaining eigenfrequencies reads:

$$C_1(\delta\omega + \Delta + \frac{a_1}{r^6}) = \frac{r^3 E_0^2}{4\hbar}(C_1 + C_2),$$

$$C_2(\delta\omega - \Delta + \frac{a_2}{r^6}) = \frac{r^3 E_0^2}{4\hbar}(C_1 + C_2),$$  (5)

and $\delta\omega \approx \Delta + \frac{a_2 - a_1}{2r^6} \pm \sqrt{(\Delta - \frac{a_2 - a_1}{2r^6})^2 + \frac{r^6 E_0^2}{16\hbar^2}}$.

To calculate the dipole force bond, we have to integrate dipole potential $U$ over 6D two-electron coordinate space with wave functions of the system:

$$U = -\int d^3 x_1 d^3 x_2 \frac{2e^2 \psi^* z_1 z_2 \psi}{r^3}.$$ Under our

assumptions $C_{3,4} \ll C_{1,2}$, therefore the final expression for bond energy is $U \approx -\frac{2\mu^2}{r^3} \operatorname{Re} C_2^* C_1$. Finally, using (5) and $|C_1|^2 + |C_1|^2 \approx 1$ one gets

$$U \approx \mp \frac{\mu^2 E_0^2}{2\hbar\sqrt{(\Delta - \frac{a_2 - a_1}{2r^6})^2 + \frac{r^6 E_0^2}{16\hbar^2}}}.$$  (3)

For one atom, surrounded by two atom in a chain, we have to multiply (3) by 2, but also, due to time averaging, this factor disappear – time averaging leads to multiplication by factor 0.5. Besides this, we neglect the transition frequency shift $\Delta$, coming from mostly the Doppler Effect, and we neglect $a_2$ for estimations. Finally, the bond energy of interaction per atom in a string of atoms is:

$$U \approx -\frac{\mu^2 E_0^2}{2\hbar\sqrt{(\frac{a_1}{2r^6})^2 + \frac{r^6 E_0^2}{16\hbar^2}}}.$$  (4)

Figure 1 shows how this potential depends on the distance $r$ between the atoms (the field is taken to be 600 MV/m, $a_1$=6.5 in atomic units (see [2]) and $\mu = \frac{2^7 \sqrt{2}}{3^5}$ in atomic units that corresponds to hydrogen 1→2 transition.

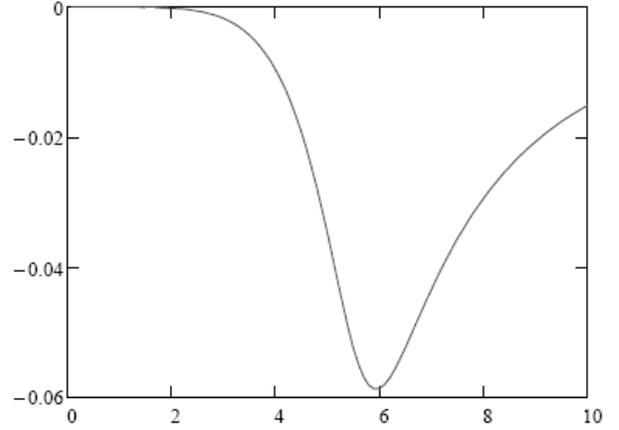

**Figure 1 Potential well.**

Figure 2 shows how minimum of the potential (in units of Kelvin, 1K ≈11000 eV) depends on the electric field amplitude (in units of 60 MV/m) for hydrogen. The hydrogen has to be cold enough to form chains of the atoms. But if one takes metals or semiconductors with much larger dipole transitions, the strings would form even for hot vapors, creating self sustained formations.

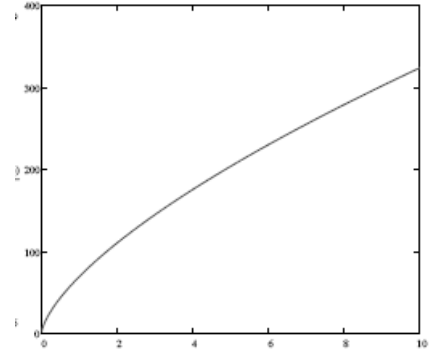

**Figure 2 Temp versuss electric field.**

In [1], some solutions for trapped fields were found for 1D and 2D cylindrical symmetries. In the simplest variant, 2D cylindrically trapped fields can form a torroid: the strings of metal (or other high $\mu$ atoms) can hold the field inside (the torroid ball lightning were reported to be seen also). The cross section of possible torroid formation is shown in Figure 3

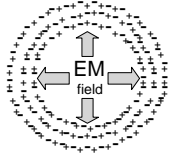

**Figure 3 Cross section of possible torroid realization of ball lightning.**

The final field-atom distribution doesn't explain, of course, the process of creation of such a formation. We think the resonant field appears in the medium in course of electric discharge and generated in a process, similar to that of the EM field generation in gas lasers.

## CONCLUSIONS

The paper describes some possible explanations of self sustained formations, similar or equivalent to the ball lightning.

## ACKNOWLEDGEMENT

Research sponsored by UT-Batelle, LLC, under contract #DE-AC05-00OR22725 for U.S. Department of Energy.